\documentclass[aps,prl,reprint,groupedaddress]{revtex4-1}
\usepackage{graphicx} 
\usepackage{dcolumn} 
\usepackage{bm}
\usepackage{hyperref}

\begin{document}

\title{ High-density symmetry energy from subthreshold hyperon production in heavy-ion collisions }
\author{Zhao-Qing Feng }
\email{Corresponding author: fengzhq@scut.edu.cn }

\affiliation{School of Physics and Optoelectronics, South China University of Technology, Guangzhou 510640, China }

\date{\today}

\begin{abstract}
The hyperon dynamics in heavy-ion collisions near threshold energy has been investigated within the quantum molecular dynamics transport model. The isospin and momentum dependent hyperon-nucleon potential and the threshold energy correction on the hyperon elementary cross section are included in the model. It is found that the high-density symmetry energy is dependent on the isospin ratios $\Sigma^{-}/\Sigma^{+}$ and $\Xi^{-}/\Xi^{0}$, in particular in the domain of high kinetic energies. The isospin diffusion in heavy-ion collisions influences the neutron/proton ratio in the high-density region. The $\Sigma^{-}/\Sigma^{+}$ ratio depends on the stiffness of symmetry energy, in particular at the beam energy below the threshold value (E$_{th}$=1.58 GeV), i.e., the kinetic energy spectra of the single ratios, excitation functions and energy spectra of the double ratios in the isotopic reactions of $^{108}$Sn + $^{112}$Sn, $^{112}$Sn + $^{112}$Sn, $^{124}$Sn + $^{124}$Sn and $^{132}$Sn + $^{124}$Sn. The double strangeness ratio $\Xi^{-}/\Xi^{0}$ weakly depends on the symmetry energy because of the hyperon-hyperon collision mainly contributing the $\Xi$ production below the threshold energy (E$_{th}$ = 3.72 GeV).

\begin{description}
\item[PACS number(s)]
 21.65.Ef, 24.10.Jv, 25.75.Dw     \\
\emph{Keywords:} High-density symmetry energy; $\Sigma^{-}/\Sigma^{+}$ ratio; $\Xi^{-}/\Xi^{0}$ ratio; Kinetic energy spectra; LQMD transport model
\end{description}
\end{abstract}

\maketitle

\section{1. Introduction}

Heavy-ion collisions provide a unique possibility for exploring the dense hadronic matter and in-medium properties of hadrons in terrestrial laboratories. Hyperons ($\Lambda, \Sigma, \Xi$ and $\Omega$) as the main ingredients of dense matter and strangeness nucleus are of importance on the properties of compact stars and hypernuclides. It is known that the appearance of hyperons in neutron stars is caused from the fermionic motion of neutron, proton and electron, e.g., the creation of $\Lambda$ through the weak process $p+e^{-} \rightarrow \Lambda+\nu_{e}$ with the chemical potential $\mu_{\Lambda}=\mu_{p}+\mu_{e^{-}}$. The threshold density with 5.3$\rho_{0}$ (the saturation density) is obtained for the $\Lambda$ appearance by neglecting the $\Lambda$-nucleon interaction. The attractive 2-body $\Lambda$-nucleon potential reduces the threshold density to be 2-4 $\rho_{0}$, softens the equation of state and leads to the maximum mass of neutron star to be 1.3-1.6$M_{\odot}$, which deviates from the recent observation with the masses of PSR J0348+0432 (2.01$\pm$ 0.04$M_{\odot}$) and J0740+6620 (2.14$^{+0.10}_{-0.09} M_{\odot}$) \cite{De10,An13,Fo16,Cr19}. One attempt to solve the hyperon puzzle in the neutron star matter is to introduce the 3-body interactions, i.e., NN$\Lambda$ \cite{Ge20}, N$\Lambda\Lambda$, NN$\Sigma$, NN$\Xi$ etc, which might be extracted from the hyperon production in heavy-ion collisions, e.g. transverse momentum spectra, collective flows etc. On the other hand, the hyperons produced in heavy-ion collisions are also sensitive probes for constraining the nuclear equation of state (EOS), which is expressed with the energy per nucleon as E($\rho,\delta$) = E($\rho,\delta=0$) + $E_{sym}(\rho)\delta^2$ + $\mathcal{O}(\delta^4)$ in terms of baryon density $\rho$ = $\rho_n$+$\rho_p$ and relative neutron excess $\delta$ = ($\rho_n - \rho_p$)/($\rho_n + \rho_p$). Both the hyperon-nucleon potential and symmetry energy influence the hyperon dynamics in heavy-ion collisions. The hyperon potential has been extensively investigated via the hyperon-nucleus scattering \cite{Ko00,Na02,Ka02}, hypernuclear properties \cite{Gi95,Ha06,Fe15,Ga16} and microscopic calculations \cite{Ha05,Fu07,Ha15,Pe16}. The isospin ratios of hyperon production may be used for extracting the high-density symmetry energy, i.e., $\Sigma^{-}/\Sigma^{+}$, $\Xi^{-}/\Xi^{0}$. Recently, the weakly repulsive $\Sigma$N potential has been argued by the Nijmegen extended-soft-core (ESC) potential model \cite{Th10} and chiral effective field theory (EFT) \cite{Pe16}. The isospin, momentum and density dependent potential influences the hyperon dynamics in heavy-ion collisions and has been investigated via transport models, i.e., collective flows, transverse momentum spectra, rapidity distribution etc \cite{Yo21,Zh21,Na22}.

The strangeness nuclear physics has been planned as one of topical issues by the large-scale scientific facilities in the world, i.e., High Intensity heavy-ion Accelerator Facility (HIAF) in China \cite{Ya19}, Alternating Gradient Synchrotron (AGS) at BNL and Relativistic Heavy-Ion Collider (RHIC-STAR) in United State \cite{Pi20,Ab22}, the LHC-ALICE at European Organization for Nuclear Research \cite{Do21}, Japan Proton Accelerator Research Complex (J-PARC) in Japan \cite{Ta22}, Dubna Nuclotron-based Ion Collider fAcility (NICA) in Russia \cite{Ko20} and so on. In theoretically, several models have been established to describe the strange particle production in heavy-ion collisions, such as the statistical thermal hadronization model \cite{An23}, intranuclear cascade model \cite{Cu24}, transport approaches based on the Boltzmann-Uehling-Uhlenbeck and quantum molecular dynamics \cite{Ha25,Bu26}. Sophisticated investigation on hyperon production in heavy-ion collisions is still expected, in particular extracting the 3-body hyperon-nucleon force and high-density symmetry energy.

In this letter, the hyperon dynamics and density dependence of symmetry energy from hyperon production in heavy-ion collisions are to be investigated with the Lanzhou quantum molecular dynamics (LQMD) model. The isospin ratios $\Sigma^{-}/\Sigma^{+}$ and $\Xi^{-}/\Xi^{0}$ in the isotopic reactions below threshold energies are systematically analyzed for constraining the high-density symmetry energy.

\section{2. Model description }

In the model, the production of resonances with the mass below 2 GeV, hyperons ($\Lambda$, $\Sigma$, $\Xi$) and mesons ($\pi$, $\eta$, $K$, $\overline{K}$, $\rho$, $\omega$) is coupled to the reaction channels via meson-baryon and baryon-baryon collisions \cite{Fe11,Fe18}. The temporal evolutions of nucleons and nucleonic resonances ($\Delta$(1232), N$^{\ast}$(1440), N$^{\ast}$(1535), etc) are described by Hamilton's equations of motion under the self-consistently generated 2-body and 3-body potentials with the well-known Skyrme effective force. The particle production, cluster and hypercluster formation in heavy-ion collisions and hadron induced reactions (proton, antiproton, meson etc) have been well described by the LQMD transport model.

The symmetry energy is composed of the kinetic energy difference from the fermionic motion of neutron and proton in nuclear matter, the local potential and the momentum interaction at the density $\rho$, which reads as \cite{Fe12}
\begin{equation}
E_{sym}(\rho)=\frac{1}{3}\frac{\hbar^{2}}{2m_{N}}\left(\frac{3}{2}\pi^{2}\rho\right)^{2/3}+E_{sym}^{loc}(\rho)+E_{sym}^{mom}(\rho)
\end{equation}
with $m_{N}$ being the nucleon mass.
The stiffness of symmetry energy is adjusted by the local part, which has the density dependent form as follows
\begin{equation}
E_{sym}^{loc}(\rho)=\frac{1}{2}C_{sym}(\rho/\rho_{0})^{\gamma_{s}},
\end{equation}
and the supersoft case
\begin{equation}
E_{sym}^{loc}(\rho)=a_{sym}(\rho/\rho_{0})+b_{sym}(\rho/\rho_{0})^{2}.
\end{equation}
The parameters $C_{sym}$, $a_{sym}$ and $b_{sym}$ are taken as the values of 52.5 MeV, 43 MeV, -16.7 MeV, respectively. The stiffness parameter $\gamma_{s}$ can be adjusted for the different density dependence of symmetry energy, e.g., 0.3, 0.5, 1 and 2 corresponding to the slope parameter $L=3\rho_{0}(\partial E_{sym}/\partial \rho)|_{\rho=\rho_{0}}$ being 42 MeV, 53 MeV, 82 MeV and 139 MeV, respectively. The supersoft symmetry energy leads to the slope value of 24 MeV. All cases cross at the saturation density with the value of 31.5 MeV. It is noticed that the symmetry energy manifests the different trend in the low-density region and in the suprasaturation density domain, e.g., the hard symmetry energy with $\gamma_{s}=$2 leading to the larger value above the normal density $\rho_{0}$, but the lower energy below $\rho_{0}$. The isospin diffusion in heavy-ion collisions is influenced by the symmetry energy, in which more repulsive interaction for neutrons in the neutron-rich matter is enforced by the larger symmetry energy. Shown in Fig. 1 is the density dependence of symmetry energy with different stiffness and the neutron/proton ratio at the suprasaturation densities with the range of $\rho>1.5\rho_{0}$ in the reaction of $^{124}$Sn + $^{124}$Sn at the incident energy of 1\emph{A} GeV. The neutron/proton ratio in the high-density region is below the average value (1.48) of reaction system and increases with the kinetic energy. The ratio is reduced with the hard symmetry energy because of more repulsive interaction for neutrons in the dense matter. The neutron-neutron and proton-proton collisions mainly contribute to the created isospin ratios, i.e., $\pi^{-}/\pi^{+}$, $K^{0}/K^{+}$, $\Sigma^{-}/\Sigma^{+}$, $\Xi^{-}/\Xi^{0}$ etc.

\begin{figure}
\includegraphics[width=8 cm]{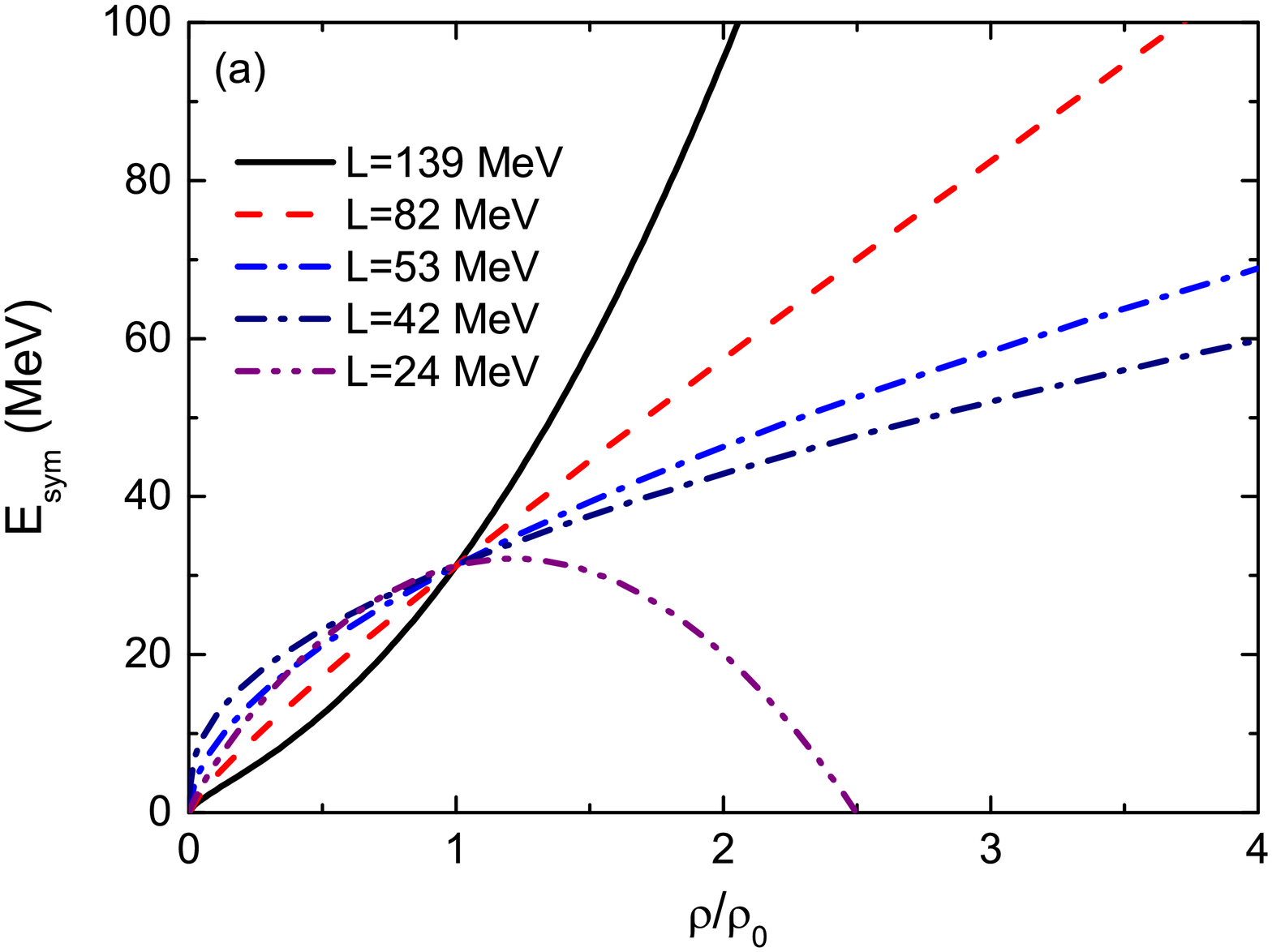}
\includegraphics[width=8 cm]{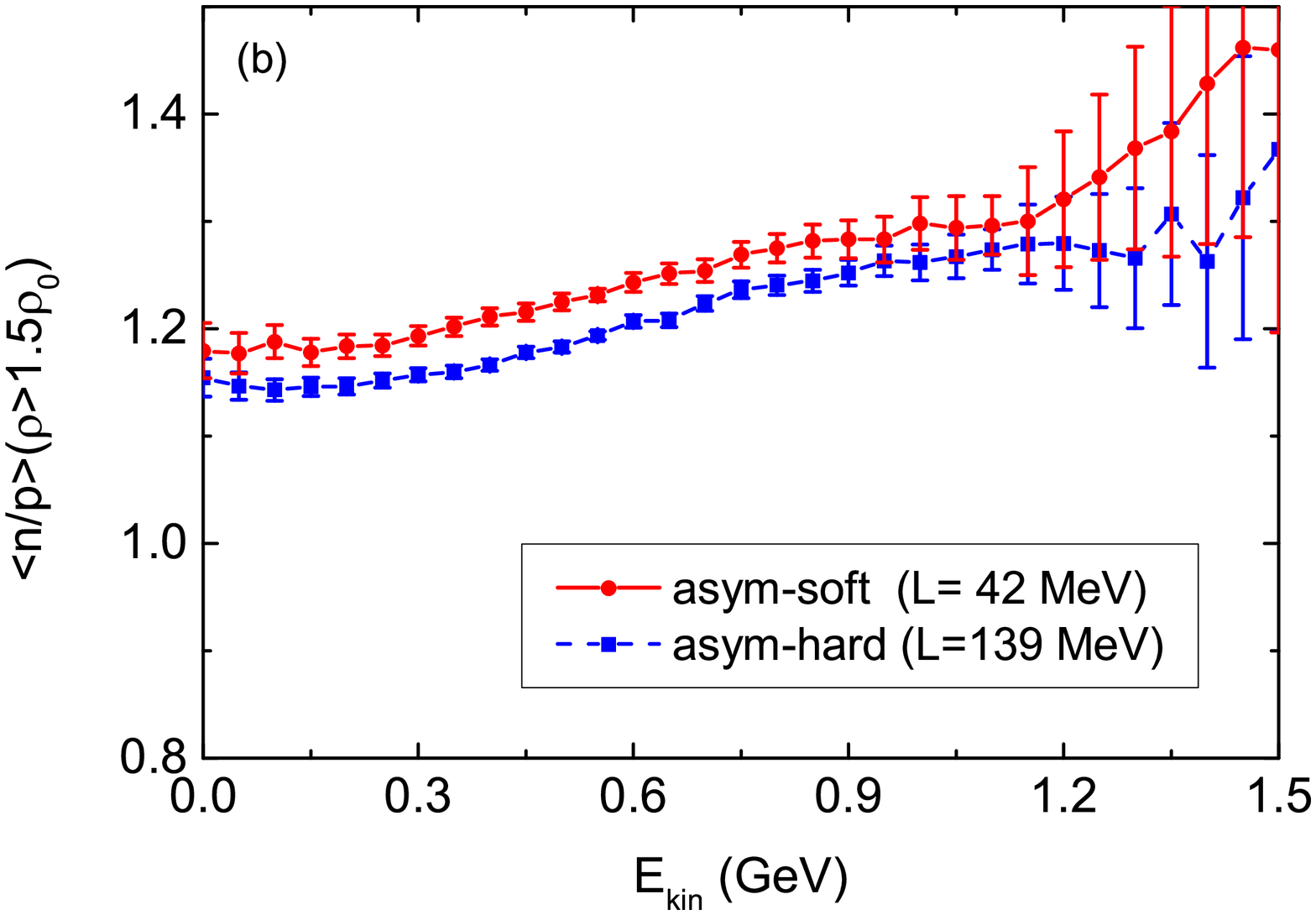}
\caption{(a) Density dependence of nuclear symmetry energy with different stiffness and (b) neutron/proton ratio within the density range $\rho>1.5\rho_{0}$ in the reaction of $^{124}$Sn+$^{124}$Sn at the incident energy of 1\emph{A} GeV. }
\end{figure}

In the LQMD transport model, the hyperons are created by the direct two, three and four-body processes via the baryon-baryon and meson-baryon collisions as follows \cite{Fe20}
\begin{eqnarray}
&& BB \rightarrow BYK,  BB \rightarrow BBK\overline{K},  B\pi(\eta) \rightarrow YK,  YK \rightarrow B\pi,     \nonumber \\
&& B\pi \rightarrow NK\overline{K}, Y\pi \rightarrow B\overline{K}, \quad  B\overline{K} \rightarrow Y\pi, \quad YN \rightarrow \overline{K}NN,  \nonumber \\
&& BB \rightarrow B\Xi KK, \overline{K}B \leftrightarrow K\Xi, YY \leftrightarrow N\Xi, \overline{K}Y \leftrightarrow \pi\Xi.
\end{eqnarray}
Here the symbols corresponding to B(N, $\triangle$, N$^{\ast}$), Y($\Lambda$, $\Sigma$), $\Xi(\Xi^{0}, \Xi^{-}$), $\pi(\pi^{-}, \pi^{0}, \pi^{+})$, K(K$^{0}$, K$^{+}$), $\overline{K}$($\overline{K}^{0}$, K$^{-}$). The elementary cross sections are parameterized by fitting the available experimental data and the Clebsch-Gordan coefficients for the isospin channels. Furthermore, the elastic scattering and strangeness-exchange reactions between strangeness and baryons have been considered through the channels of $KB \rightarrow KB$, $YB \rightarrow YB$ and $\overline{K}B \rightarrow \overline{K}B$ parameterized in Ref. \cite{Cu90}. The charge-exchange reactions between the $KN \rightarrow KN$ and $YN \rightarrow YN$ channels are included by using the same cross sections with the elastic scattering, such as $K^{0}p\rightarrow K^{+}n$, $K^{+}n\rightarrow K^{0}p$ etc \cite{Fe13}.

The evolution of hyperons is also determined by the Hamiltonian, which is composed of the Coulomb potential and strong interaction as
\begin{eqnarray}
H_{Y}=\sum^{N_{Y}}_{i=1}(V^{Coul}_{i}+V^{Y}_{opt}\left(\bm{p}_{i},\rho_{i})+\sqrt{\bm{p}^{2}_{i}+m_{Y}^{2}}\right).
\end{eqnarray}
Here the $N_{Y}$ is the total number of hyperons ($\Lambda$, $\Sigma$, $\Xi$). The Coulomb potential $V^{Coul}_{i}$ is estimated by the point-charge interaction between the hyperon and charged baryons. The optical potentials of $\Lambda$ and $\Xi$ are calculated on the basis of the light-quark counting rule, which result in the strength values at the saturation density being -32 MeV and -16 MeV, respectively. The $\Sigma$N optical potential is estimated by
\begin{eqnarray}
V^{\Sigma}_{opt}(\bm{p}_{i},\rho_{i}) = &&  V_{0}(\rho_{i}/\rho_{0})^{\gamma_{s}}+V_{1}(\rho_{n} - \rho_{p}) t_{\Sigma}\rho_{i}^{\gamma_{s}-1}/\rho_{0}^{\gamma_{s}}        \nonumber\\
&& + C_{mom}\ln(\epsilon\bm{p}^{2}_{i}+1).
\end{eqnarray}
Here, the isospin quantities are taken as $\bm{t}_{\Sigma}$=1, 0, and -1 for $\Sigma^{-}$, $\Sigma^{0}$ and $\Sigma^{+}$, respectively. The $C_{mom}$ and $\epsilon$ are 1.76 MeV and 500 $c^{2}/$GeV$^{2}$. The values of the isoscalar $V_{0}=$14.8 MeV and isovector $V_{1}=$67.8 MeV are obtained by fitting the calculated results from the next-to-leading order (NLO) in chiral EFT with a cutoff value of 600 MeV \cite{Ha15}. The hyperon dynamics in heavy-ion collisions is influenced by the optical potential. The effective mass $m^{\ast}_{Y}=V^{Y}_{opt}(\textbf{p}_{i}=0,\rho_{i})+m_{Y}$ is implemented into the threshold energy correction for the hyperon production, e.g. $\sqrt{s^{\ast}}=m_{B}+m^{\ast}_{Y}+m^{\ast}_{K}$ for the channel $BB \rightarrow BYK$,  The repulsive potential will enhance the threshold energy and consequently result in the reduction of hyperon production. However, the attractive potential manifests an opposite contribution.

\section{3. Results and discussion}

The isospin observables of strange particles in heavy-ion collisions are influenced by the nucleon-nucleon collisions and the in-medium effect. More neutron-neutron collision enhances the production of K$^{0}$, $\Sigma^{-}$ and $\Xi^{-}$, but associated with the density dependence of symmetry energy. Therefore, the ingredient of neutron and proton number in the high-density region is of significance for the isospin ratio to extract the symmetry energy information. Shown Fig. 2 is the kinetic energy spectra of K$^{0}$/K$^{+}$ and $\Sigma^{-}/\Sigma^{+}$ produced in the high-density region for the reaction of $^{124}$Sn + $^{124}$Sn at 1.5\emph{A} GeV. The isospin ratios K$^{0}$/K$^{+}$ and $\Sigma^{-}/\Sigma^{+}$ are estimated from the temporal evolution of reaction system. It is obvious that the soft symmetry energy with L=42 MeV leads to the enhancement of the isospin ratios, which is caused from the larger neutron-neutron collision probability for the soft case in the density region of $\rho>\rho_{0}$. The invariant energy spectra are also analyzed for $\Sigma$ and $\Xi$ production in collisions of $^{124}$Sn+$^{124}$Sn at 1.5\emph{A} GeV and 2.5\emph{A} GeV in Fig. 3, respectively. The difference of $\Sigma^{-}$ and $\Sigma^{+}$ yields is mainly caused from the hadron-hadron collisions, mean-field potential, symmetry energy and in-medium correction on the elementary cross section. The $\Sigma^{-}$ potential is more repulsive in comparison with the one of $\Sigma^{+}$ in the neutron-rich matter. The production of $\Xi^{-}$ and $\Xi^{0}$ is also influenced by the four factors. But the same optical potentials are used for $\Xi^{-}$ and $\Xi^{0}$ evolution. More investigation on the $\Xi$-nucleon interaction from the microscopic theory is still needed, in particular including the 3-body force, isospin and momentum dependent form.

\begin{figure*}
\includegraphics[width=16 cm]{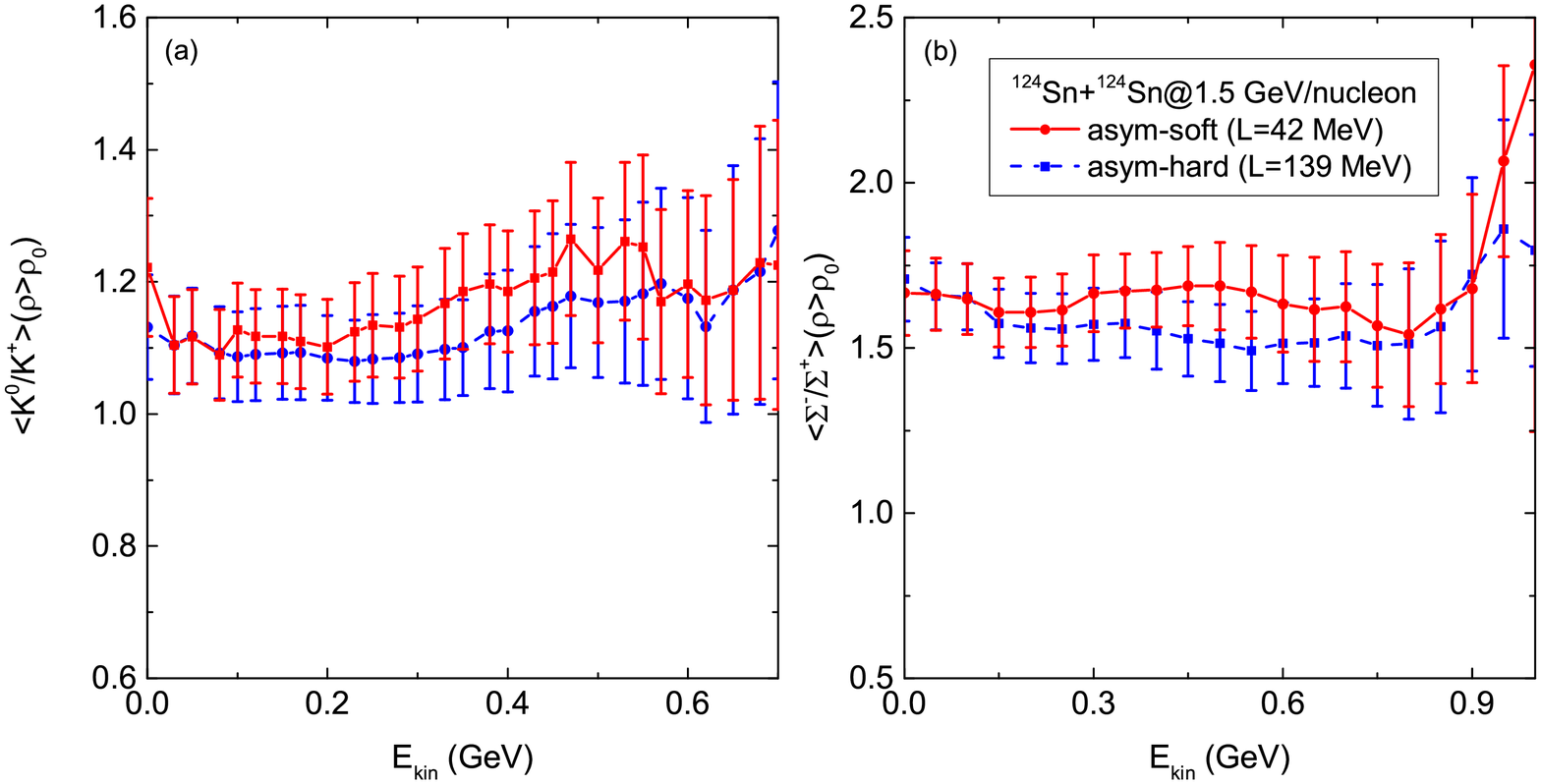}
\caption{ Kinetic energy distributions of K$^{0}$/K$^{+}$ and $\Sigma^{-}/\Sigma^{+}$ produced in the high-density range of $\rho>\rho_{0}$ for the reaction of $^{124}$Sn + $^{124}$Sn at the incident energy of 1.5 GeV/nucleon, respectively. }
\end{figure*}

\begin{figure*}
\includegraphics[width=16 cm]{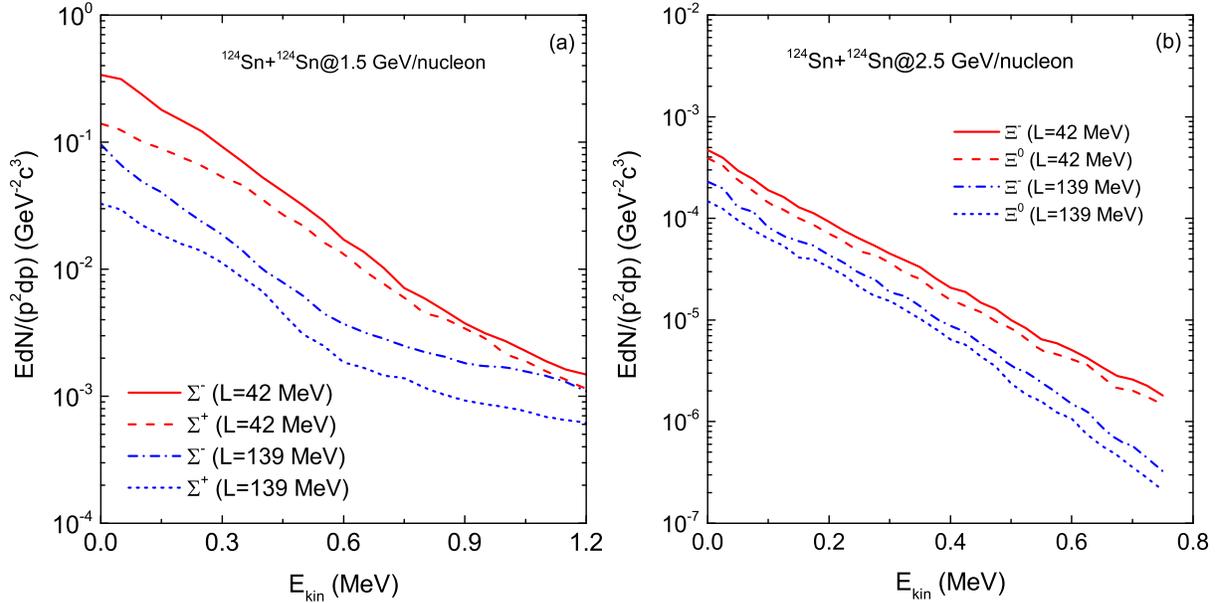}
\caption{ Comparison of inclusive spectra of $\Sigma$ and $\Xi$ produced in collisions of $^{124}$Sn + $^{124}$Sn near threshold energies with the soft and hard symmetry energy, respectively. }
\end{figure*}

The high-density symmetry energy is very important for the mass-radius relation and maximum mass of the neutron star, the frequency of gravitational wave by the binary neutron star merging. The hyperon might appear above the 2$\rho_{0}$ baryon density because the nucleon Fermi energy become high enough to create hyperons. Shown in Fig. 4 is a comparison of the kinetic energy spectra for $\Sigma^{-}/\Sigma^{+}$ and $\Xi^{-}/\Xi^{0}$ in the reaction of $^{124}$Sn+$^{124}$Sn below the threshold energy. It is obvious that the $\Sigma^{-}/\Sigma^{+}$ ratio is sensitive to the stiffness of symmetry energy. However, the $\Xi^{-}/\Xi^{0}$ ratio weakly depends on the symmetry energy owing to the $\Xi$ production mainly contributed from the secondary collisions, i.e., the reaction channels of $\overline{K}N \rightarrow K\Xi$, $\overline{K}Y \rightarrow \pi\Xi$ and $YY \rightarrow N\Xi$. The effect becomes more obvious with increasing the N/Z ratio of reaction system. More experiments are expected for measuring the kinetic energy or momentum spectra in the near future.

\begin{figure*}
\includegraphics[width=16 cm]{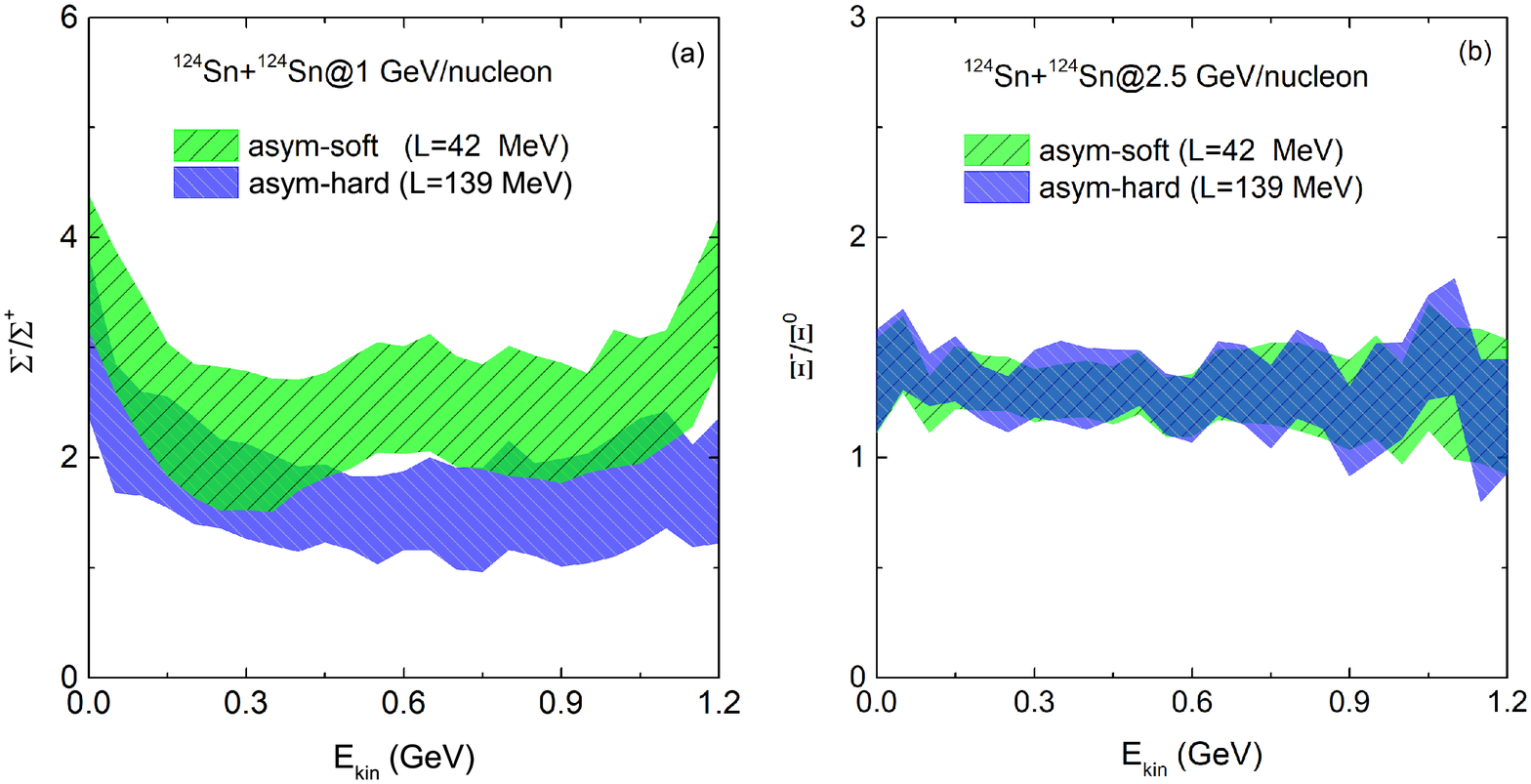}
\caption{ Kinetic energy distributions of $\Sigma^{-}/\Sigma^{+}$ (left panel) and $\Xi^{-}/\Xi^{0}$ (right panel) in the reaction of $^{124}$Sn+$^{124}$Sn with different stiffness of symmetry energy. }
\end{figure*}

The dense hadronic matter formed in heavy-ion collisions is associated with the collision centrality, beam energy and reaction system. The isospin ratio of hyperons for extracting the high-density symmetry energy might be influenced by the Coulomb interaction, rescattering process, hyperon-nucleon potential, elementary cross section for hyperon production in hadron-hadron collisions. The incident energy dependence of the hyperon multiplicities and isospin ratios in the reaction of $^{124}$Sn+$^{124}$Sn is calculated as shown in Fig. 5. The hyperon production rapidly increases with the beam energy. To eliminate the Coulomb interaction, we also analyzed the double ratios of $\Sigma^{-}/\Sigma^{+}$ and $\Xi^{-}/\Xi^{0}$ from a comparison of the systems $^{132}$Sn + $^{124}$Sn/$^{108}$Sn + $^{112}$Sn and $^{124}$Sn + $^{124}$Sn/$^{112}$Sn + $^{112}$Sn. It is featured that the double ratio (DR) of $\Sigma^{-}/\Sigma^{+}$ is sensitive to the high-density symmetry energy, in particular in the regime of high kinetic energy. The soft symmetry energy with L=42 MeV leads to the larger DRs owing to the more $\Sigma^{-}$ production from the neutron-neutron collisions in dense matter. The DRs of $\Xi^{-}/\Xi^{0}$ manifests the horizontal distribution with the DR value of neutron/proton of reaction system. The $\Xi$ production is mainly contributed from the secondary collisions, which reduces the isospin effect of $\Xi^{-}/\Xi^{0}$. It should be noticed that the production of $\Xi$ in heavy-ion collisions and its capture by a nuclear fragment are helpful for understanding the hyperon puzzle in neutron star and for creating the double-strangeness hypernucleus, respectively. The 3-body interactions of NN$\Lambda$, N$\Lambda$$\Lambda$ and NN$\Xi$ might be extracted from the hyperon production in heavy-ion collisions and the investigation of hypernuclear properties.

\begin{figure*}
	\includegraphics[width=16 cm]{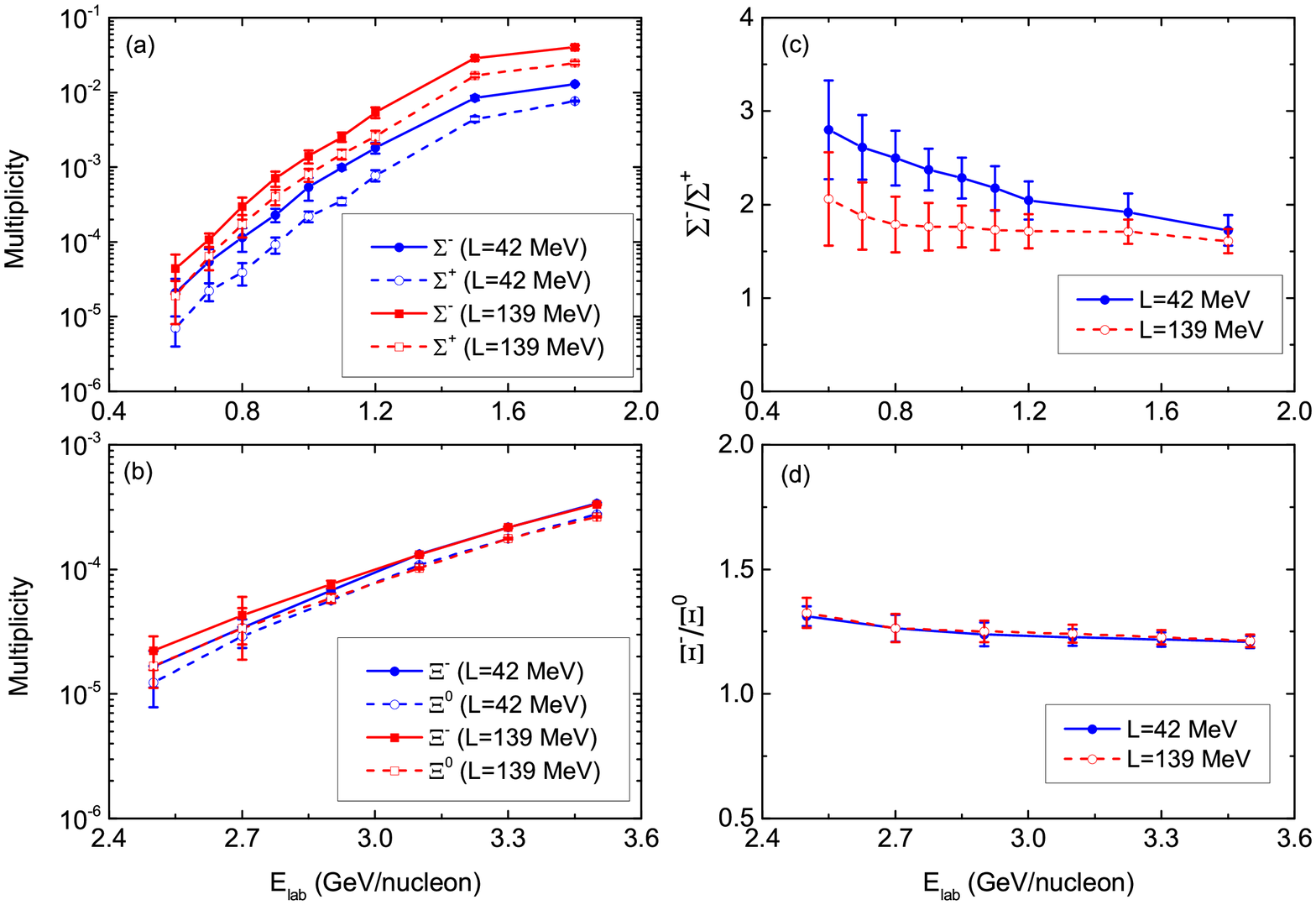}
	\caption{Excitation functions of the total multiplicities of $\Sigma^{-}$, $\Sigma^{+}$, $\Xi^{-}$ and $\Xi^{0}$ in collisions of $^{124}$Sn+$^{124}$Sn with the different stiffness of symmetry energy. }
\end{figure*}

\begin{figure*}
	\includegraphics[width=16 cm]{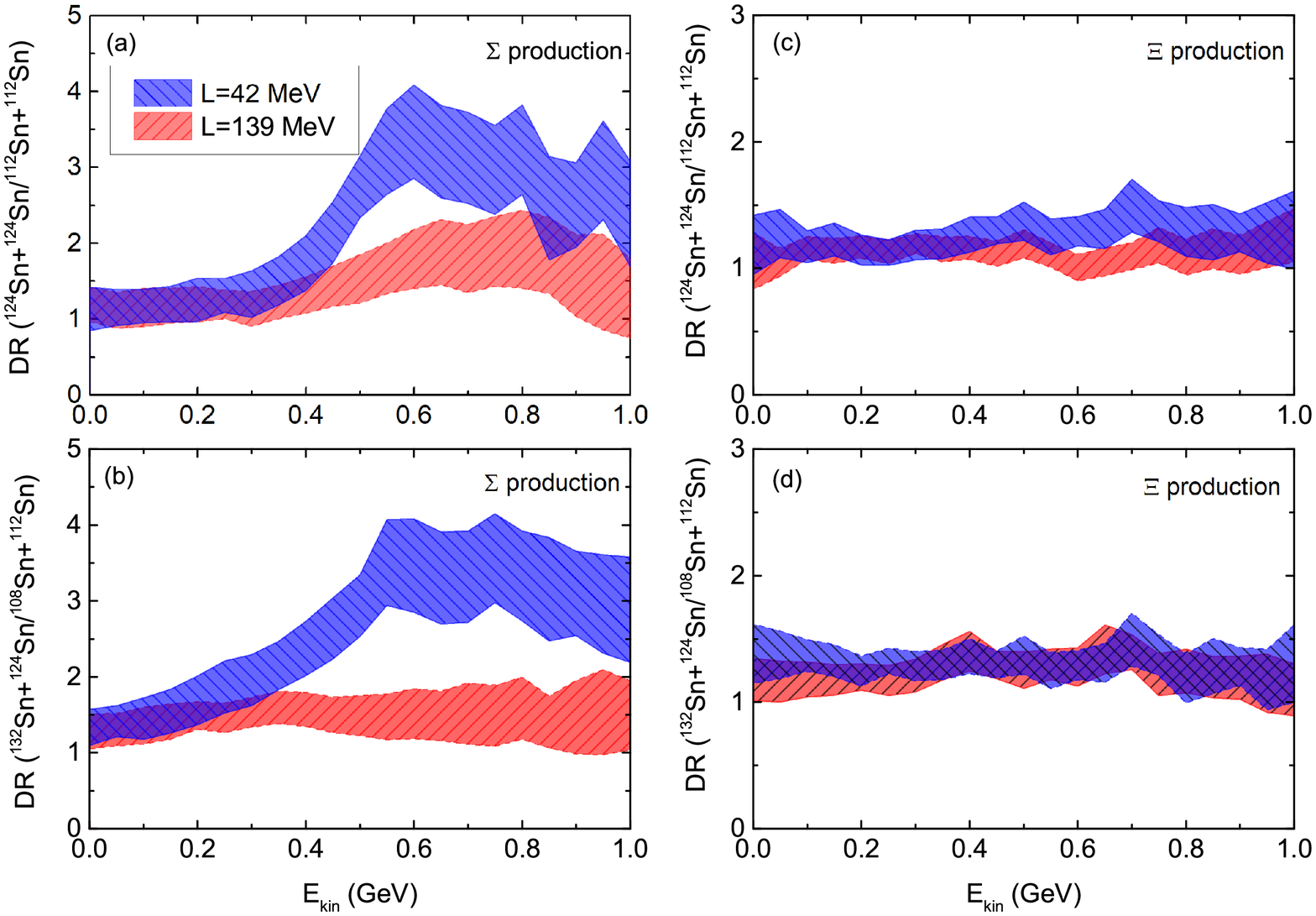}
	\caption{ Kinetic energy spectra of the double ratios of $\Sigma^{-}/\Sigma^{+}$ at 1\emph{A} GeV and $\Xi^{-}/\Xi^{0}$ at 2.5\emph{A} GeV in the isotopic systems of $^{108}$Sn + $^{112}$Sn, $^{112}$Sn + $^{112}$Sn, $^{124}$Sn + $^{124}$Sn and $^{132}$Sn + $^{124}$Sn. }
\end{figure*}

\section{4. Conclusions}

In summary, the hyperon dynamics in heavy-ion collisions below the threshold energy and the high-density symmetry energy from the isospin ratios of strange particles are systematically investigated by the LQMD transport model. The neutron/proton ratio at the suprasaturation density is influenced by the isospin diffusion and associated with the high-density symmetry energy. The isospin ratios K$^{0}$/K$^{+}$ and $\Sigma^{-}/\Sigma^{+}$ in the high-density region are enhanced with softening the stiffness of symmetry energy. The energy spectra of hyperons are related to the hyperon-nucleon potential and the symmetry energy. The $\Sigma^{-}/\Sigma^{+}$ ratio at the subthreshold energy is sensitive to the stiffness of the high-density symmetry energy, in particular in the domain of high kinetic energies. The energy spectra of $\Xi^{-}/\Xi^{0}$ are weakly dependent on the symmetry energy. The double $\Sigma^{-}/\Sigma^{+}$ ratios in the isotopic reactions of $^{132}$Sn + $^{124}$Sn/$^{108}$Sn + $^{112}$Sn and $^{124}$Sn + $^{124}$Sn/$^{112}$Sn + $^{112}$Sn at the beam energy of 1\emph{A} GeV are promising observables for constraining the high-density symmetry energy. Further experiments are expected in the near future, i.e., the strangeness physics at HIAF, NICA etc.

\textbf{Acknowledgements}
This work was supported by the National Natural Science Foundation of China (Projects No. 12175072 and No. 11722546) and the Talent Program of South China University of Technology (Projects No. 20210115).

\textbf{Declaration of Competing Interest}
The author declares that they have no known competing financial interests or personal relationships that could have appeared to influence the work reported in this paper.

\textbf{Data availability}
No data was used for the research described in the article.


\begin{thebibliography}{99}

\bibitem{De10} P. Demorest, T. Pennucci, S. Ransom, M. Roberts, and J. Hessels, Nature \textbf{467}, 1081 (2010).
\bibitem{An13} J. Antoniadis \emph{et al.}, Science \textbf{340}, 6131 (2013).
\bibitem{Fo16} E. Fonseca \emph{et al.}, Astrophys. J. \textbf{832}, 167 (2016).
\bibitem{Cr19} H. T. Cromartie \emph{et al.}, Nat. Astron. \textbf{4}, 72 (2019).
\bibitem{Ge20} D. Gerstung, N. Kaiser, and W. Weise, Eur. Phys. J. A \textbf{56}, 175 (2020).
\bibitem{Ko00} Y. Kondo, J.K.Ahn, H.Akikawa, \emph{et al.}, Nucl. Phys. A \textbf{676}, 371 (2000).
\bibitem{Na02} J. Nagata, H. Yoshino, V. Limkaisang, \emph{et al.}, Phys. Rev. C \textbf{66}, 061001(R) (2002).
\bibitem{Ka02} T. Kadowaki,  J. Asai, W. Imoto, \emph{et al.}, Eur. Phys. J. A \textbf{15}, 295 (2002).
\bibitem{Gi95} B.E. Gibson and E.V. Hungerford III, Phys. Rep. \textbf{257}, 349 (1995).
\bibitem{Ha06} O. Hashimoto and H. Tamura, Prog. Part. Nucl. Phys. \textbf{57}, 564 (2006).
\bibitem{Fe15} A. Feliciello and T. Nagae, Rep. Prog. Phys. \textbf{78}, 096301 (2015).
\bibitem{Ga16} A. Gal, E. V. Hungerford, and D. J. Millener, Rev. Mod. Phys. \textbf{88}, 035004 (2016).
\bibitem{Ha05} J. Haidenbauer and Ulf-G. Mei{\ss}ner, Phys. Rev. C \textbf{72}, 044005 (2005).
\bibitem{Fu07} Y. Fujiwara, Y. Suzuki, and C. Nakamoto, Prog. Part. Nucl. Phys. \textbf{58}, 439 (2007).
\bibitem{Ha15} J. Haidenbauer and Ulf-G. Mei{\ss}ner,  Nucl. Phys. A \textbf{936}, 29 (2015).
\bibitem{Pe16} S. Petschauer, J. Haidenbauer, N. Kaiser \emph{et al.},  Eur. Phys. J. A \textbf{52}, 15 (2016).
\bibitem{Th10} Th. A. Rijken, M. M. Nagels, and Y. Yamamoto, Nucl. Phys. A \textbf{835}, 160 (2010).
\bibitem{Yo21} G. C. Yong, Z. G. Xiao, Y. Gao, and Z. W. Lin, Phys. Lett. B \textbf{820}, 136521 (2021).
\bibitem{Zh21} D. C. Zhang, H. G. Cheng, and Z. Q. Feng, Chin. Phys. Lett. \textbf{38}, 092501 (2021).
\bibitem{Na22} Y. Nara, A. Jinno, K. Murase, and A. Ohnishi, Phys. Rev. C \textbf{106}, 044902 (2022).
\bibitem{Ya19} J. C. Yang, J. W. Xia, G. Q. Xiao, \emph{et al.}, Nucl. Instr. Meth. B \textbf{317}, 263 (2013).
\bibitem{Pi20} C. Pinkenburg, N. N. Ajitanand, J. M. Alexander, \emph{et al.}, Phys. Rev. Lett. \textbf{2743}, 68 (1999).
\bibitem{Ab22} M. S. Abdallah \emph{et al.}, STAR Collaboration, Phys. Rev. Lett. \textbf{128}, 202301 (2022).
\bibitem{Do21} B. D\"{o}nigus, Nucl. Phys. A \textbf{904-905}, 547c (2013).
\bibitem{Ta22} H. Tamura, Prog. Theor. Exp. Phys. \textbf{2012}, 02B012 (2012).
\bibitem{Ko20} V. I. Kolesnikov, V. D. Kekelidze, V. A. Matveev \emph{et al.}, Physica Scripta, \textbf{95}(9), 094001 (2020).
\bibitem{An23} A. Andronic, P. Braun-Munzinger, J. Stachel, \emph{et al.}, Phys. Lett. B \textbf{697}, 203 (2011).
\bibitem{Cu24} J. Cugnon, P. Deneye, J. Vandermeulen, Nucl. Phys. A \textbf{500}, 701 (1989).
\bibitem{Ha25} C. Hartnack, H. Oeschler, Y. Leifels, E. Bratkovskaya, and J. Aichelin, Phys. Rep. \textbf{510}, 119 (2012).
\bibitem{Bu26} O. Buss \emph{et al.}, Phys. Rep. \textbf{512}, 1 (2012).
\bibitem{Fe11} Z. Q. Feng, Phys. Rev. C \textbf{84}, 024610 (2011).
\bibitem{Fe18} Z. Q. Feng, Nucl. Sci. Tech. \textbf{29}, 40 (2018).
\bibitem{Fe12} Z. Q. Feng, Phys. Rev. C \textbf{85}, 014604 (2012).
\bibitem{Fe20} Z. Q. Feng, Phys. Rev. C \textbf{102}, 044604 (2020).
\bibitem{Cu90} J. Cugnon, P. Deneye, J. Vandermeulen, Phys. Rev. C \textbf{41}, 1701 (1990).
\bibitem{Fe13} Z. Q. Feng, Nucl. Phys. A \textbf{919}, 32 (2013).

\end{thebibliography}
\end{document}